# A Thermodynamic Hypothesis Regarding Optimality Principles for Flow Processes in Geosystems

HUI-HAI LIU

Earth Sciences Division
Lawrence Berkeley National Laboratory
Berkeley, CA 94720, USA
hhliu@lbl.gov

**Abstract**

This paper proposes a new thermodynamic hypothesis that states that a nonlinear natural system that is not isolated and involves positive feedbacks tends to minimize its resistance to the flow process through it that is imposed by its environment. We demonstrate that the hypothesis is consistent with flow behavior in saturated and unsaturated porous media, river basins, and the Earth-atmosphere system**.** While optimization for flow processes has been previously discussed by a number of researchers in the literature, the unique contribution of this work is to indicate that only the driving process is subject to optimality when multiple flow processes are simultaneously involved in a system.

1. **Introduction**

   Optimality principles refer to that state of a physical system is controlled by an optimal condition that is subject to physical and/or resource constrains. Optimality principles have been used, as a holistic approach, for flow processes in several important geosystems.

   *Rodriguez-Iturbe et al* [1992] postulated the minimization of energy expenditure rate (MEE) principles at both local and global scales for channel networks in a river basin. *Rinaldo et al.* [1992] developed modeling approaches using MEE to generate optimal channel networks, and compared their results with those from natural river basins. Striking similarity was observed for natural and optimal networks in their fractal aggregation structures and other relevant features. Using the similar optimality principle, *Liu* [2011a] developed a group of (partial differential) governing equations for steady-state optimal landscapes (including both channel networks and associated hillslopes) using calculus of variations. Most recently, *Liu* [2011b] applied the MEE to unsaturated media involving gravity-dominated unstable flow, a common phenomenon. He found that unsaturated hydraulic conductivity for water flow, in this case, is not only a function of water saturation or capillary pressure (as indicated in the classic theory by *Buckingham* [1907]), but also a power function of water flux. Furthermore, he showed that conductivities in both the unsaturated flow in soils and water flow in landscapes follow similar power-function relationships (with water flux), but with different exponent values.

   The maximum entropy production (MEP) principle, initially proposed by *Paitridge* [1975], has been shown to be useful for predicting behavior of the Earth-atmosphere system. The MEP principle states that a flow system subject to various flows or gradients will tend towards a steady-state position of the maximum thermodynamic entropy production [*Nieven*, 2010].

   The physical origin of these optimality principles is not totally clear at this point. They cannot be directly deduced from the currently existing thermodynamic laws [*Bejan*,

2010], although some effort has been made to make the connections with these laws [e.g., *Martyushev and Seleznev*, 2006; *Sonnino and Evslin*, 2007]. *Bejan* [2000] argued that these thermodynamics laws deal largely with processes within black-boxes and were not developed for describing flow structures for flow processes (within these boxes) that are associated with the optimality principles. We agree with this assessment. As an effort to develop a general principle to resolve the above issue, *Bejan* [2000] proposed the constructal law that states that "for a finite-size open system to persist in time (to survive) it must evolve in such a way that it proves easier and easier access to the currents that flow through it". The constructal law has been used to explain many natural phenomena including the related optimality principles.

However, a geosystem generally involves a number of flow processes and empirical evidence indicates that not all the flow processes are subject to optimality, as will be discussed below. A general principle to identify which flow process is subject to optimality is still lacking in the literature. The objective of this short communication is to propose a thermodynamic hypothesis for identifying the optimal process. The usefulness of our proposal is also demonstrated by its consistency with several flow systems.

## 2. A thermodynamic hypothesis

We propose the following thermodynamic hypothesis: *a nonlinear natural system that is not isolated and involves positive feedbacks tends to minimize its resistance to the flow process through it that is imposed by its environment.*

It is important to emphasize that we deal with a system here that is not isolated and subject to mass and/or energy transfer between it and its environment. The system must be nonlinear and involve positive feedbacks such that the relevant flow process plays an important role in forming the flow paths and, at the same time, the formation of these flow paths would further enhance the flow. In such a system, several flow processes (including heat flow [transfer]) may co-exist. For example, a river basin at least involves surface water flow and soil erosion processes associated with water flow. Our hypothesis indicates that the system (river basin) does not tend to provide minimum resistance to all the flow processes (e.g., soil erosion [soil particle flow]), but to the process *imposed by its environment* (atmosphere [rainfall] and upstream), or the driving process that is water flow. We will further demonstrate this point in the next section with different geo-systems.

Our hypothesis is motivated by the notion that natural systems tend to evolve to more and more uniformity [e.g., *Bejan*, 2000]. A classic example of this is an isolated system in which all the physical properties are completely uniform when it becomes equilibrium. In a system that is not isolated, because of energy and/or mass exchange with its environment, it never becomes completely uniform spatially, but approaches another kind of uniformity that we call "dynamic uniformity" here. Take a river basin as an example. Water entering the river basin has larger energy than the downstream water. Thus, the former has the tendency to equilibrate with the latter. To do so, the former tends to flow to the latter location as quickly as possible and with as little energy loss as possible. Note that the less energy loss means better uniformity (in term of energy). In this regard, the dynamic uniformity corresponds to the MEE. Therefore, our hypothesis is equivalent to that "*a nonlinear natural system that is not isolated and involves positive feedbacks tends to maximize the degree of dynamic uniformity for the flow process imposed by its environment.*" To some extent, the concept of dynamic uniformity is similar to entropy in an isolated system in which entropy tends to be maximized.

Our hypothesis also implies the existence of often observed hierarchy structures in some natural systems. This simply because it applies to both the whole system, such as, for example, a river basin, and a subsystem, such as a part of a river basin associated with a branch of a river network. In this case, the minimum resistance to water flow should occur in the entire river basin and a part of it. As a result, similar flow structures (river/channel geometry) occur at different levels of spatial scales.

Our hypothesis is consistent with the constructal law of *Bejan* [2000] that states that "for a finite-size open system to persist in time (to survive) it must evolve in such a way that it proves easier and easier access to the currents that flow through it". The constructal law, proposed as a general physical law, can be used to explain many natural phenomena [e.g., *Bejan*, 2000]. Our hypothesis is more specific regarding the conditions under which the optimality occurs for flow processes in geosystems. It is indicated that not all the systems are subject to optimization process except for the nonlinear ones that are not isolated and involve positive feedbacks. Especially, for a system involving multiple flow processes, only the driving process is subject to resistance minimization.

### 3. Discussions

In this section, we demonstrate that our hypothesis is consistent with flow behavior in several typical geo-systems.

### 3.1 **Water flow in saturated porous media**

Steady-state saturated flow in porous media, a linear system, is employed here to demonstrate that the minimization of flow resistance, in general, may not occur for a linear system, as implied by our hypothesis.

The energy expenditure rate, $EE$, for water flow in saturated porous media across an entire flow domain $\Omega$ can be given by [*Liu*, 2011]

$$EE = \int_\Omega K(\nabla h)^2 dV \qquad (1)$$

where $K$ is saturated hydraulic conductivity, $h$ is hydraulic head, and $V$ is porous-medium volume. Note that the local energy expenditure rate, $K(\nabla h)^2$, is equal to the product of Darcy flux and hydraulic gradient. The local entropy production rate will be the local energy expenditure rate divided by the temperature [*Li*, 2003]. Thus, as previously indicated, under isothermal conditions, the minimization of energy expenditure rate is identical to the minimization of entropy production rate.

By minimizing $EE$ (Eq. (1)) using Euler equation [e.g., *Weinstock*, 1974], we obtain:

$$\nabla \bullet \nabla h = 0 \qquad (2)$$

Comparing the above equation with the following continuity equation for water flow in saturated porous media:

$$\nabla \bullet (K\nabla h) = 0 \qquad (3)$$

we can conclude that the flow process satisfies the MEE only for homogeneous porous media, because (2) is the same as (3) for constant $K$. In fact, this result can also be directly obtained from the well-known *Prigogine*'s [1955] theorem of minimization of entropy production which is valid for linear irreversible processes. However, when $K$ is a function of space, which is generally the case for real-world problems, steady-state flow processes will not follow the MEE principle, because Eq. (2) [or Eq. (1)] is not identical to (3) any more. An alternative argument is that for linear systems, there are no feedback mechanisms because, for example, $K$ distributions in saturated porous media are not impacted by water flow at all.

From the above discussion, it is evident that not all flow processes follow optimality principles in terms of flow resistance, especially for linear systems. That is why our hypothesis excludes linear systems.

3.2 **Water flow in unsaturated porous media**

In this subsection, we use unsaturated flow in porous media to show that for a nonlinear system, the existence of positive feedback mechanisms, as indicated by our hypothesis, is required for generating the minimum resistance to water flow.

For a downward, gravity-dominated water flow in unsaturated porous media, fingering (unstable) flow patterns generally occur. In such a system, the flow process imposed by its environment is water flow because water intrudes into porous media. It is well-known that unsaturated flow is a nonlinear process; the conductivity for water flow is a function of properties associated with water flow itself. Furthermore, positive feedback mechanisms occur here. When fingering takes place, more water tends to flow through fingers and therefore makes fingers grow, because fingers provide flow paths with smaller flow resistance compared with uniform flow. Thus, our hypothesis indicates that the MEE should apply here.

Based on the MEE, *Liu* [2011b] derived the following relationship for gravity-dominated unsaturated flow:

$$K_{unsat} = F(h_c)\left(\frac{|q|}{K}\right)^a \qquad (4)$$

where $K_{unsat}$ is unsaturated hydraulic conductivity, $F(h_c)$ is a function of capillary pressure head, $h_c$, $|q|$ is the magnitude of Darcy flux, and $a$ is a constant whose value seems to be about 0.5 from a limited number of previous studies [*Liu*, 2011b]. The above relationship is supported by experimental observations and is consistent with some empirical models [*Liu*, 2011b].

We emphasize that one grand challenge facing us in the area of hydrogeology is the need to develop physical laws for large-scale multiphase-flow problems. At a local scale, fluid distribution is mainly controlled by capillarity and is not sensitive to flow conditions. That is why relative permeability at a local scale can be successfully described as a function of saturation (or capillary pressure) only [*Buckingham*, 1907]. At a large scale, this is not longer the case, although local-scale relationships have been widely used at large scales because alternatives are unavailable. Our result (Eq. (4)) suggests that functional forms of large-scale relationships to describe multiphase flow are very likely different from their counterparts at local scales, which cannot be resolved from upscaling based on the same functional forms as those at local scales. It is our hope that the optimality approach may provide an important way to obtain such large-scale relationships.

However, MEE and Eq. (4) cannot be applied to nonlinear unsaturated flow in an upward direction, because for such a flow system, there is no a positive feedback and therefore no fingering either. Flow structures (such as fingering) are generally a signature of positive feedback mechanisms. Without these mechanisms, small disturbance to water flow that always exists in nature would not grow into flow structures, but rather die out during the flow process. As indicated in our hypothesis, the existence of positive feedback mechanisms is required for the flow resistance to be minimized.

### 3.3 Flow processes in a river basin and the Earth-climate system

Finally, we show that our hypothesis allows for a possible reconciliation of the MEE and the MEP optimality principles, with a focus on the importance of the dominant process imposed by the environment.

The consistency between water flow in a river basin and our hypothesis was given above in Section 2. In this open and nonlinear system, positive feedback exists. During rain water flows down hill, water tends to flow more in channels/rivers to make them grow once they are initialized by water flow process, because resistance to flow process in these structures is much smaller than that to flow along a hill surface. Also, as previously indicated, water flow is the dominant process imposed by its environment; all other processes involved in a river basin (such as soil erosion) are initialized and driven by water flow. Thus, based on our hypothesis, the MEE applies to water flow in a river basin.

Flow processes in the Earth-atmosphere system are more complex. Some inconsistency seems to exist between the MEE and the MEP, even they have been successfully applied in different geosystems. Under isothermal conditions, energy expenditure rate is proportional to entropy production rate [*Li*, 2003]. Thus, under such conditions, the MEP requires that a river network forms in such a way that the energy expenditure rate for water flow should be at its maximum, which directly contradicts the MEE supported by empirical data for a river basin.

To clarify this issue, we must recognize that the Earth-atmosphere system generates minimum flow resistance to the (active) "flow process imposed by its environment", but not for other (reactive) processes that just help with the resistance minimization for the former process, as indicated by our hypothesis. The Earth receives radiation from the hot Sun and transfers the received heat into space. The atmosphere and oceans act as a fluid system that transports heat from the hot region to the cold one with general circulation [*Ozawa et al.*, 2003], because the convection process is more efficient in transferring heat than the conduction process [*Bejan*, 2000]. Obviously, in this system, the "flow process imposed by its environment" is the heat flow; the heat flow process is the initiator for other flow processes.

Under steady-state flow conditions, the average heat flow rate is closely related to entropy production in the Earth-atmosphere system, as shown by *Paltridge* [1978] and *Ozawa et al.* [2003] and the MEP corresponds to the maximum convective heat transport [*Paltridge*, 1978]. The latter was further confirmed by *Clausse et al.* [2012] who showed that the temperature distributions on the Earth surface is consistent with a principle (derived from the constructal law) that convective heat flow rate from equator region to the pole region is maximized, or the resistance to heat flow is minimized at a global scale. In this case, the MEP happens to be a by-product of this heat-flow optimization process. Along this line, the MEP in the Earth-atmosphere system and the MEE in a river basin are all consistent with our hypothesis and can be unified in terms of minimizing resistance to the "flow process imposed by its environment", or the driving process.

The "maximum heat transfer" hypothesis was initially proposed in a theoretical study of the well-known Rayleigh-Benard convection by *Malkus and Veronis* [1954]. They used that hypothesis to identify wave lengths for the stable convection cells. However, their work has been criticized on the ground that their predicted relationship between the wave length and Rayleigh number is not always consistent with experimental observations [*Koschmieder*, 1993]. While the mathematical technique used by *Markus and Veronis* [1954] is not very reliable for nonlinear problems [*Getling*, 1998], there have been no observations that directly disprove "the maximum heat transfer" hypothesis itself [*Koschmieder*, 1993].

The above discussion has been based on an important argument that the minimum heat-flow resistance results in the maximum heat flow rate and the MEP in the Earth-atmosphere system. This argument can be further justified as follows. Under steady-state flow conditions, the entropy production in the Earth-atmosphere system is given by *Paltridge* [1978] and *Ozawa et al.* [2003]:

$$S' = \int_A \frac{dQ}{T} = Q\left(\frac{1}{T_L} - \frac{1}{T_H}\right) = Q\frac{\Delta T}{T_{av}^2[1-\left(\frac{\Delta T}{T_{av}}\right)^2]} \tag{5}$$

where $S'$ is the entropy production rate, $A$ is the boundary (Earth surface), $Q$ is heat flow through the boundary, $T_L$ and $T_H$ are average Earth surface temperatures in the low and high temperature regions, respectively, $T_{av}$ is the average temperature in both high and low temperature regions, and $\Delta T = T_H - T_L$. (Heat flows from the high-temperature region to the low temperature region.) The above equation represents the fact that the entropy production by some processes associated with turbulence is completely discharged into the surrounding system through the boundary under steady-state conditions [*Ozawa et al.*, 2003].

Since $(\Delta T/T_{av})^2$ is generally on the order of 1-2% and $T_{av}$ can be reasonably determined based on global solar heat current and Earth surface radiation into space only, Eq. (5) can be rewritten as

$$S' \approx Q\frac{\Delta T}{T_{av}^2} \tag{6}$$

where $T_{av}$ is considered a well constrained parameter that does not depend on $Q$ and $\Delta T$. *Clausse et al.* [2013] indicated that convective flow rate from high temperature region to the low temperature region can be expressed as

$$Q = C(\Delta T)^{3/2} \tag{7}$$

where C is a constant related to Earth radius and properties of fluid involved in heat flow. Accordingly, the conductance for heat flow will be:

$$K_H = \frac{Q}{\Delta T} = C(\Delta T)^{1/2} \tag{8}$$

Based on Eqs. (6), (7) and (8), we have

$$S' = \frac{Q^{5/3}}{T_{av} C^{2/3}} \tag{9}$$

$$K_H = C^{2/3} Q^{1/3} \tag{10}$$

Clearly, Eqs. (9) and (10) show that the maximum $K_H$ (or the minimum heat flow resistance) corresponds to the maximum heat flow rate ($Q$) and entropy production rate ($S'$). It is also of interest to note that the formulation for the heat conductance (Eq. (10)) is consistent with the conductivity relations for water flow in river basins and unsaturated porous media, derived from the optimality principle [*Liu*, 2011a; 2011b], because they all are power functions of flux (or flow rate) with positive exponent values.

## 4. Conclusion

Optimality principles have been used, as a holistic approach, for flow processes in several important geosystems. We proposed a new thermodynamic hypothesis that states that a nonlinear natural system that is not isolated and involves positive feedbacks tends to minimize its resistance to the flow process through it that is imposed by its environment. In the other worlds, only the driving process is subject to optimality when multiple flow processes are involved. We also demonstrate that the hypothesis is consistent with water flow behavior in saturated and unsaturated porous media.

**Acknowledgement**

We are indebted to Drs. Lianchong Li and Dan Hawkes at Lawrence Berkeley National Laboratory for their critical and careful review of a preliminary version of this manuscript. This work was supported by the U.S. Department of Energy (DOE), under DOE Contract No. DE-AC02-05CH11231.